\documentclass[11pt]{article}
\pdfoutput=1
\usepackage{graphicx}
\usepackage{amssymb,amsmath}
\usepackage{multirow}
\usepackage{cite,color,url}
\usepackage[colorlinks=true
,urlcolor=blue
,anchorcolor=blue
,citecolor=blue
,filecolor=blue
,linkcolor=blue
,menucolor=blue
,linktocpage=true
,pdfproducer=medialab
,pdfa=true
]{hyperref}

\usepackage{epsfig,psfrag,rotating,soul}
\usepackage{rotfloat}

\evensidemargin \oddsidemargin
\allowdisplaybreaks
\def\C{\tilde C_0}
\def\dk{d\tilde k}
\def\D{\big(k^2-m_1^2\big) \big((k+p_1)^2-m_2^2\big) \big((k+p_1+p_2)^2-m_3^2\big)}
\def\e{\epsilon}

\begin{document}
\thispagestyle{empty}

\def\thefootnote{\fnsymbol{footnote}}

\begin{flushright}
IFT-UAM/CSIC-14-027
\end{flushright}

\vspace{0.5cm}

\begin{center}

{\large\sc {\bf  Remark on the one-loop $\mathbf{Z}$ form factors for LFV $\mathbf{Z}$-penguin diagrams in SUSY}}

\vspace*{.7cm} 

{\sc
E. Arganda$^{1}$%
\footnote{email: {\tt \href{mailto:ernesto.arganda@unizar.es}{ernesto.arganda@unizar.es}}}%
, M.J. Herrero$^{2}$%
\footnote{email: {\tt \href{mailto:maria.herrero@uam.es}{maria.herrero@uam.es}}}%
}

\vspace*{.7cm}

{\sl
$^1$Departamento de F\'isica Te\'orica, Facultad de Ciencias,\\
Universidad de Zaragoza, E-50009 Zaragoza, Spain 

\vspace*{0.1cm}

$^2$Departamento de F\'isica Te\'orica and Instituto de F\'isica Te\'orica,
UAM/CSIC\\
Universidad Aut\'onoma de Madrid, Cantoblanco, 28049 Madrid, Spain
 
}

\end{center}

\vspace*{0.1cm}

\begin{abstract}
\noindent
In this note we shortly comment on our analytical results in PRD73(2006)055003
for the $Z$ boson one-loop form factors contributing to the Lepton Flavour Violating $Z$ penguin diagrams in SUSY. In a recent communication [arXiv:1312.5318v1] it has been pointed out a mistake in our formulas for the chargino contribution to the $Z$-form factor, $F_L^{(c)}$, and these authors have included corrections to our results in a way that we do not agree with.  We wish to clarify here what are the correct results for these form factors.  
 
\end{abstract}

\def\thefootnote{\arabic{footnote}}
\setcounter{page}{0}
\setcounter{footnote}{0}

\newpage

In a recent communication \cite{Krauss:2013gya}, the authors are quoting our paper \cite{Arganda:2005ji}
  claiming that it contains a mistake in the formulas of the form factors describing the effective Lepton Flavour Violating $l_i-l_j-Z$ vertex that are generated by the one-loop diagrams with MSSM charginos-sneutrinos and neutralinos-sleptons. We wish to clarify in the following our disagreements/agreements with their corrections and set our corrected formulas for the one-loop $Z$ form factors.   

Our convention here for the form factors $F_{L,R}$ defining the effective $Z_\mu l_j l_i$ 
vertex, the name used for all the functions, couplings, indexes, parameters and particle labels are as in \cite{Arganda:2005ji}. Thus, the $F_{L,R}$ form factors for the LFV vertex $Z l_j l_i$ are introduced as
\begin{equation}
-i\gamma_\mu\left[F_L P_L+F_R P_R\right]\,\,\,,\, {\rm with}\,\,,\,\, P_{R,L}=\frac{1}{2}(1\pm \gamma_5) \,.  
\end{equation}
\begin{figure}[hbtp]
  \begin{center} 
    \includegraphics[width=150mm]{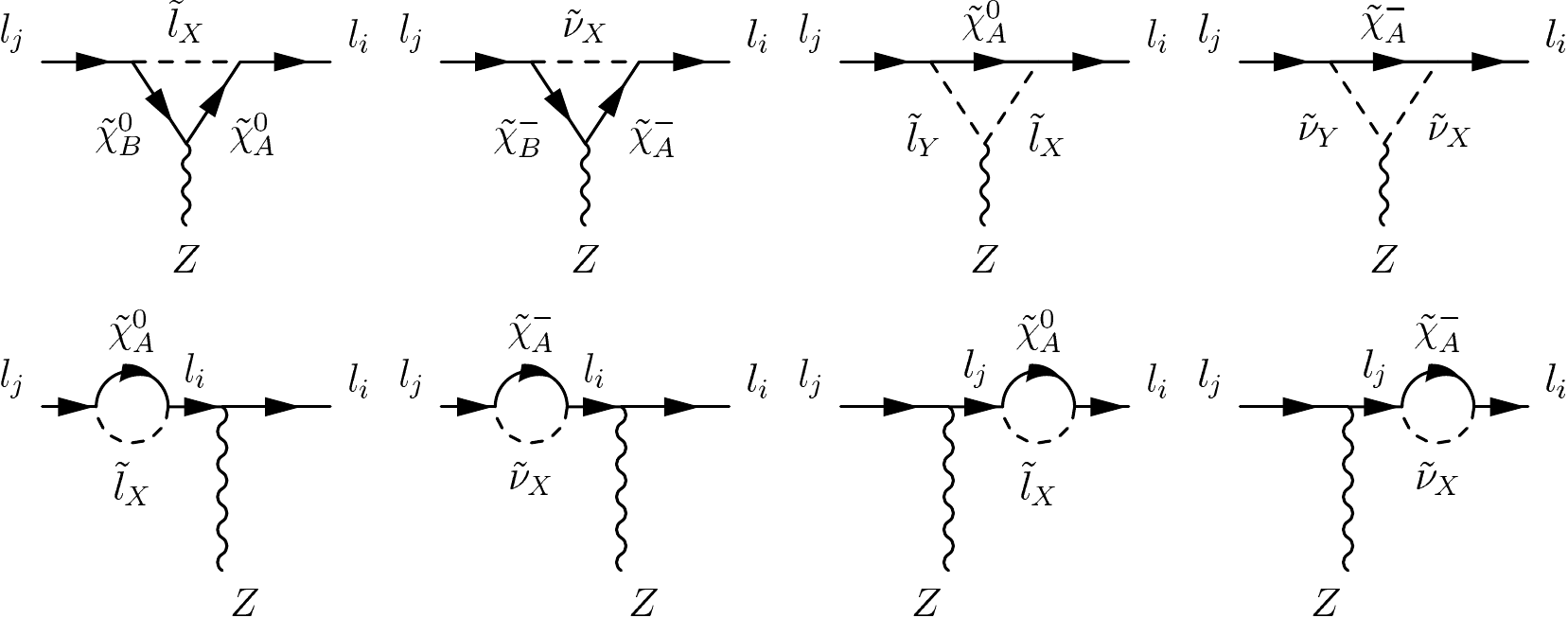}
    \caption{Relevant SUSY one-loop diagrams for the $Z$-mediated contributions to 
    LFV processes. The particle content is as in the MSSM.
    }\label{Z_diagrams} 
  \end{center}
\end{figure}

The contributing SUSY one-loop diagrams are those included in \cite{Arganda:2005ji} that, for clearness,  are also collected here in Fig.~\ref{Z_diagrams}. Notice that we are concerned with the contributing loops in the MSSM context, therefore containing only the sparticles of the MSSM: 
sneutrinos (SUSY partners of $\nu_L$), sleptons, charginos and neutralinos. We are not concerned here with the possibility of including additional sneutrinos (nor additional neutrinos) beyond the MSSM particle content, like the SUSY partners of the $\nu_R$ which have been considered in \cite{Krauss:2013gya} and also in the recent reply to \cite{Krauss:2013gya}, that just appeared in
\cite{Ilakovac:2014ypa} where they also comment on this same issue.
 
Our results for the $Z$-boson form factors in \cite{Arganda:2005ji} that included the two contributions, from neutralinos
$(n)$ and charginos $(c)$, which we repeat here for clearness, were
\begin{equation}\label{Zformfactors}
F_{L(R)} = F_{L(R)}^{(n)} + F_{L(R)}^{(c)} \,,
\end{equation}
with
\begin{eqnarray}\label{Zformfactors_uncorrected}
F_L^{(n)} &=& -\frac{1}{16 \pi^2} \left\{ N_{iBX}^R N_{jAX}^{R \ast} \left[  E_{BA}^{R(n)} 2 C_{24}(m_{\tilde{l}_X}^2, m_{\tilde{\chi}_A^0}^2, m_{\tilde{\chi}_B^0}^2) - E_{BA}^{L(n)} m_{\tilde{\chi}_A^0} m_{\tilde{\chi}_B^0} C_0(m_{\tilde{l}_X}^2, m_{\tilde{\chi}_A^0}^2, m_{\tilde{\chi}_B^0}^2) \right] \right. \nonumber \\
&+& \left. N_{iAX}^R N_{jAY}^{R \ast} \left[ 2 Q_{XY}^{\tilde{l}} C_{24}(m_{\tilde{\chi}_A^0}^2, m_{\tilde{l}_X}^2, m_{\tilde{l}_Y}^2) \right] + N_{iAX}^R N_{jAX}^{R \ast} \left[ Z_L^{(l)} B_1(m_{\tilde{\chi}_A^0}^2, m_{\tilde{l}_X}^2) \right] \right\} \,, \\
F_R^{(n)} &=& \left. F_L^{(n)} \right|_{L \leftrightarrow R} \,, \\
F_L^{(c)} &=& -\frac{1}{16 \pi^2} \left\{ C_{iBX}^R C_{jAX}^{R \ast} \left[  E_{BA}^{R(c)} 2 C_{24}(m_{\tilde{\nu}_X}^2, m_{\tilde{\chi}_A^-}^2, m_{\tilde{\chi}_B^-}^2) - E_{BA}^{L(c)} m_{\tilde{\chi}_A^-} m_{\tilde{\chi}_B^-} C_0(m_{\tilde{\nu}_X}^2, m_{\tilde{\chi}_A^-}^2, m_{\tilde{\chi}_B^-}^2) \right] \right. \nonumber \\
&+& \left. C_{iAX}^R C_{jAY}^{R \ast} \left[ 2 Q_{XY}^{\tilde{\nu}} C_{24}(m_{\tilde{\chi}_A^-}^2, m_{\tilde{\nu}_X}^2, m_{\tilde{\nu}_Y}^2) \right] + C_{iAX}^R C_{jAX}^{R \ast} \left[ Z_L^{(l)} B_1(m_{\tilde{\chi}_A^-}^2, m_{\tilde{\nu}_X}^2) \right] \right\} \,, \\
F_R^{(c)} &=& \left. F_L^{(c)} \right|_{L \leftrightarrow R} \,,
\end{eqnarray}
where the indices are $A,B=1,..,4$, $X,Y=1,..,6$ in the contributions from the
neutralino sector and $A,B=1,2$, $X,Y=1,2,3$ in the contributions from the chargino
sector, and a summation over the various indices is understood.

In these formulas we were (as we are now) using a standard notation for the one-loop integrals, $B_0$, $B_1$, $C_0$, $C_{24}$,  based on the original ones in \cite{Passarino:1978jh}. Specifically, these integrals are defined by
\begin{eqnarray}
\frac i{16\pi^2} B_0(p_1^2,m_1^2,m_2^2)&=& \int\dk \frac{1}{\big(k^2-m_1^2\big) \big((k+p_1)^2-m_2^2\big)} \,, \\ 
\frac i{16\pi^2} B_\mu(p_1^2,m_1^2,m_2^2)&=&\int\dk \frac{k_\mu}{\big(k^2-m_1^2\big) \big((k+p_1)^2-m_2^2\big)} \,, \\
\frac i{16\pi^2} C_0(p_1^2,p_2^2,m_1^2,m_2^2,m_ 3^2)&=&\int\dk\frac{1}{\D} \,, \\
 \frac i{16\pi^2} C_{\mu\nu}(p_1^2,p_2^2,m_1^2,m_2^2,m_ 3^2)&=&\int\dk\frac{k_\mu k_\nu}{\D} \,,
\end{eqnarray}
and by
\begin{eqnarray}
B_\mu &=& p_{1\mu}B_1 \,, \\
C_{\mu\nu}&=&p_{1\mu}p_{1\nu}C_{21}+p_{2\mu}p_{2\nu}C_{22}+
(p_{1\mu}p_{2\nu}+p_{2\mu}p_{1\nu})C_{23}+g_{\mu\nu} C_{24} \,,
\end{eqnarray}
where we have used a short notation for the integrals in $D=4-\e$ dimension,
\begin{eqnarray}
\dk &\equiv &\mu^{4-D}\frac{d^D k}{(2\pi)^D} \,.
\end{eqnarray}
We have repeated here again this same computation and we have found a mistake in eqs.(3) and (5) where there is a missing finite term. The corresponding equations with our corrected results are as follows
\begin{eqnarray}\label{Zformfactors_corrected}
F_L^{(n)} &=& -\frac{1}{16 \pi^2} \left\{ N_{iBX}^R N_{jAX}^{R \ast} \left[  E_{BA}^{R(n)} (2 C_{24}(m_{\tilde{l}_X}^2, m_{\tilde{\chi}_A^0}^2, m_{\tilde{\chi}_B^0}^2)-\frac{1}{2}) - E_{BA}^{L(n)} m_{\tilde{\chi}_A^0} m_{\tilde{\chi}_B^0} C_0(m_{\tilde{l}_X}^2, m_{\tilde{\chi}_A^0}^2, m_{\tilde{\chi}_B^0}^2) \right] \right. \nonumber \\
&+& \left. N_{iAX}^R N_{jAY}^{R \ast} \left[ 2 Q_{XY}^{\tilde{l}} C_{24}(m_{\tilde{\chi}_A^0}^2, m_{\tilde{l}_X}^2, m_{\tilde{l}_Y}^2) \right] + N_{iAX}^R N_{jAX}^{R \ast} \left[ Z_L^{(l)} B_1(m_{\tilde{\chi}_A^0}^2, m_{\tilde{l}_X}^2) \right] \right\} \,, \\
F_R^{(n)} &=& \left. F_L^{(n)} \right|_{L \leftrightarrow R} \,, \\
F_L^{(c)} &=& -\frac{1}{16 \pi^2} \left\{ C_{iBX}^R C_{jAX}^{R \ast} \left[  E_{BA}^{R(c)} (2 C_{24}(m_{\tilde{\nu}_X}^2, m_{\tilde{\chi}_A^-}^2, m_{\tilde{\chi}_B^-}^2)-\frac{1}{2}) - E_{BA}^{L(c)} m_{\tilde{\chi}_A^-} m_{\tilde{\chi}_B^-} C_0(m_{\tilde{\nu}_X}^2, m_{\tilde{\chi}_A^-}^2, m_{\tilde{\chi}_B^-}^2) \right] \right. \nonumber \\
&+& \left. C_{iAX}^R C_{jAY}^{R \ast} \left[ 2 Q_{XY}^{\tilde{\nu}} C_{24}(m_{\tilde{\chi}_A^-}^2, m_{\tilde{\nu}_X}^2, m_{\tilde{\nu}_Y}^2) \right] + C_{iAX}^R C_{jAX}^{R \ast} \left[ Z_L^{(l)} B_1(m_{\tilde{\chi}_A^-}^2, m_{\tilde{\nu}_X}^2) \right] \right\} \,, \\
F_R^{(c)} &=& \left. F_L^{(c)} \right|_{L \leftrightarrow R} \,,
\end{eqnarray}
Thus, with this missing term now added, our corrected analytical results for $F_L^{(c)}$ do agree with those revised in \cite{Krauss:2013gya}. However, we disagree in the way they corrected our results, and this we would like to comment next.
Specifically, these authors say the following statement that we do not agree with : '...whereas $C_{24}$ is defined by the authors of \cite{Arganda:2004bz} as $4C_{24}(m_0^2,m_1^2,m_2^2)=B_0(m_1^2,m_2^2)+m_0^2C_0(m_0^2,m_1^2,m_2^2)$...'. However in our work \cite{Arganda:2004bz}, which is about the different subject 'Lepton flavor violating Higgs boson decays from massive seesaw neutrinos' we did not write this equation, in fact the function $C_{24}$ did not appear at all in this work. Our analytical results in \cite{Arganda:2004bz} were written, in contrast, in terms of a different function, $\C$ which is defined in our works \cite{Arganda:2004bz} and \cite{Arganda:2005ji} by
\begin{eqnarray}
\frac i{16\pi^2} \C(p_1^2,p_2^2,m_1^2,m_2^2,m_ 3^2)&\equiv& \int\dk\frac{k^2}{\D} \,,
\end{eqnarray}
which is related to $B_0$ and $C_0$ by
\begin{eqnarray}
\C(p_1^2,p_2^2,m_1^2,m_2^2,m_ 3^2)=B_0(p_2^2,m_2^2,m_3^2) + m_1^2 C_0 (p_1^2,p_2^2,m_1^2,m_2^2,m_ 3^2) \,,
\end{eqnarray}
and whose finite part at zero external momenta was given in  \cite{Arganda:2005ji} as
\begin{eqnarray}
\C(m_1^2, m_2^2, m_3^2) &=& 1 - \frac{1}{m_2^2 - m_3^2} \left( \frac{m_1^4 \log{m_1^2} - m_2^4 \log{m_2^2}}{m_1^2 - m_2^2} - \frac{m_1^4 \log{m_1^2} - m_3^4 \log{m_3^2}}{m_1^2 - m_3^2} \right) \,.
\end{eqnarray}
We want to emphasise that our results in \cite{Arganda:2004bz} are correct.  
We would also like to remark that we did not mention at all in any of these two works the alternative function $C_{00}$, contrary to what is transmitted in \cite{Krauss:2013gya} and in  \cite{Ilakovac:2014ypa}. Therefore, their comments on the relations of our results with $C_{00}$ are misleading. Moreover, we would also like to clarify  that in order to make a proper comparison between our results here and those in terms of $C_{00}$, one should make the identification of $C_{24}$ with $C_{00}$.

Finally, we would also like to correct here the formula (B10) of the Appendix B in \cite{Arganda:2005ji}. 
We wrote incorrectly the finite part of $C_{24}$ in terms of the finite part of 
$\C$. Thus, our equation (B10) in \cite{Arganda:2005ji}, where it was incorrectly set $ C_{24}= \frac{1}{4}\C$, should be replaced by this now corrected (B10):
\begin{eqnarray}
C_{24}(m_1^2, m_2^2, m_3^2)=\frac{1}{4}\C(m_1^2, m_2^2, m_3^2)+\frac{1}{8} \,.
\end{eqnarray}
We have redone all the plots in \cite{Arganda:2005ji} with the corrected formulas  
and we have found no difference at all in any of the numerical results and plots of this article. Thus, these corrections are  numerically totally irrelevant for our work in \cite{Arganda:2005ji}. We have also checked that these corrections do not change any of our numerical results in refs.\cite{Arganda:2007jw} and \cite{Arganda:2008jj} where the $Z$-penguin diagrams were also participating.
Finally, as an extra check of our formulas for the one-loop $Z$-form factors, we have also found agreement with the corresponding LFV $Zl_il_j$ effective vertices of ref.\cite{Illana:2002tg}. In contrast we do not find agreement with the formulas for the $Z$-penguin form factors in ref.\cite{Fukuyama:2005bh}. 
 
\section*{Acknowledgements}

We warmly thank Xabier Marcano, for helping us with all the checking process, and we also thank C\'edric Weiland for discussions about the above mentioned missing term in a previous version of this note.
E.~A. is financially supported by the Spanish DGIID-DGA grant 2013-E24/2 and the Spanish MICINN grants FPA2012-35453 and CPAN-CSD2007-00042. 
The work of M.~J.~H. is partially supported by the European Union FP7 ITN
INVISIBLES (Marie Curie Actions, PITN- GA-2011- 289442), by the CICYT through the
project FPA2012-31880, by the CM (Comunidad Autonoma de Madrid) through the project HEPHACOS S2009/ESP-1473, by the Spanish Consolider-Ingenio 2010 Programme CPAN (CSD2007-00042) and by the Spanish MINECO's "Centro de Excelencia Severo Ochoa" Programme under grant SEV-2012-0249.


\end{document}